\documentclass[article,prb,preprint,showpacs,preprintnumbers,amsmath,amssymb]{revtex4}
\usepackage{graphicx}
\usepackage{amsmath}

\begin{document}

\title{ \bf Simulations on the electromechanical poling of ferroelectric ceramics}
\author{\bf  Yin-Zhong Wu and Yong-Mei Tao}
\affiliation{ Jiangsu Laboratory of Advanced Functional materials,
and Department of Physics, Changshu Institute of Technology,
Changshu 215500, P.~R.~China\footnote{Email: yzwu@cslg.edu.cn}}

\begin{abstract}
Based on the two-step-switching model, the process of
electromechanical poling of a ferroelectric ceramics is simulated.
A difference of the remnant polarizations between two poling
protocols (mechanical stress is applied before and after the
application of poling field) is found from our simulations, which
is also observed in experiment. An explanation is given to
illustrate why the remnant polarization for the case that
mechanical stress is loaded after the application of electric
field is larger than the case that mechanical stress is loaded
before the application of electric field. Our simulation results
supply a proof for the validity of the two-step-switching model in
the electromechanical poling of polycrystalline ferroelectric
ceramics.
\end{abstract}


\maketitle
\section{Introduction}
Ceramic Lead-zirconate-titanate[PbZr$_{x}$Ti$_{1-x}$O$_{3}]$ is
one of the technologically most important polar oxide materials.
It may be used for ferroelectric random access memories,
high-storage dynamic RAM capacitors, sensors and actuators in
smart structures. Most of these applications require a high
remnant polarization, and an external load is applied on the
unpolable ceramics to acquire a large polarization in the poling
direction\cite{Zhou1}. On poling, the domains in the ceramics
reorient themselves to give a polarization closer to the poling
direction. Domain switching and rotation in PZT ceramics can be
triggered by either a mechanical stress or an electric field. As
is well known, a strong electric field polarizes the ceramics by
aligning the polarization of the domains as closely as possible
with the electric field. For a strong applied compressive stress
paralleling to the polarization, the domain will be rotated to
make the polarization as close as possible to $90^{0}$ from the
direction of the applied stress. If the external electric and
mechanical loads acted on the domain at the same time, a criterion
for domain switching was first proposed by Hwang~\cite{Hwang1} on
the basis of the total work done by electric and mechanical loads
for switching the domain as the driving force, which is expressed
by
\begin{equation}
\sigma_{ij}\Delta e^{s}_{ij}+E_{i}\Delta P^{s}_{i}\geq
2P_{s}E_{c},
\end{equation}
where $E_{i}$ and $\sigma_{ij}$ are the components of the applied
electric field and stress, respectively, $P^{s}_{i}$ and
$e^{s}_{ij}$ are the components of the spontaneous polarization
and spontaneous strain, respectively. However, Hwang's criterion
cannot explain the dependence of the mechanical stress on coercive
electric field, and the dependence of electric field on coercive
mechanical stress, and there is no energetic difference between
$90^{0}$ domain-switching and $180^{0}$ domain-switching. Last
year, a new criterion for domain-switching under the application
of both electric and mechanical loads in ferroelectric materials
was proposed~\cite{two-step1, two-step2}, in which the $180^{0}$
domain switching is replaced by two successive $90^{0}$
switchings. It is found that this two-step switching criterion can
model the hysteresis loops and butterfly-shape curves of
ferroelectric materials and accurately predict the coercive
electric fields under different applied mechanical stresses. In
most experimental and theoretical studies, the uniaxial
compressive stress and the applied electric field are loaded in
the same direction~\cite{Zhou1,Zhou2}. Recently, the
electromechanical poling of a ferroelectric PZT rod is
reported~\cite{Exp}. The set-up is shown in Fig. 1.
\begin{figure}
   \includegraphics[width=3.0in]{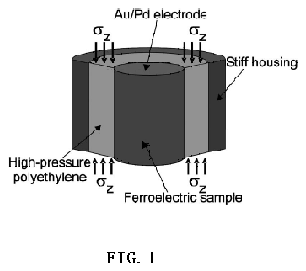}
\end{figure}

The commercial PZT PIC 151 sample is placed inside a high pressure
polyethylene tube which transfers axial compressive stress into
uniform radial compressive stress. It is shown that the
application of mechanical stress perpendicular to the electric
poling direction drastically improves the ferro- and piezoelectric
properties, and decreases the magnitude of electric field required
for poling. They clearly observe a path dependence of the
polarization buildup. The remnant polarization and the
piezoelectric coefficient of the ceramics will be different if the
mechanical stress is applied before and after the application of
electric field. And the energetic approach mentioned above could
not give an explanation on this path dependent poling result. In
this paper, we employ the two successive $90^{0}$ switchings model
to simulate the poling behavior of ferroelectric ceramics under a
radial pressures, as shown in Fig.$~1$. It is found from our
simulations that a larger remnant polarization is obtained under
the case that the mechanical load is applied after the electric
load, and this supply a new evidence for the validity of the two
successive $90^{0}$ switchings model in explanation of domain
switching behavior of the ferroelectric ceramics under the
electromechanical load.
\section{Model and Simulations}
The unpolable ferroelectric ceramics is assumed to be made up of
many randomly oriented grains, each grain is ferroelectric
tetragonal and single domain. The  90$^{0}$ switching for the
tetragonal symmetry is used as a general representative of the
non-180$^{0}$ domain switching in this paper. For a homogeneous
ceramics, the Reuss approximation~\cite{Reuss} is adopted when the
material is subjected to the applied electric field and stress.
The orientation of the grain is described by a global Cartesian
coordinate, and a local Cartesian coordinate system is also set up
along with the edges of the tetragonal cell. The orientation of
each domain can be described by three angles $\psi$, $\theta$, and
$\varphi$. The orthogonal transformation between the global
coordinate $\vec{x}$ and the local coordinate $\vec{x'}$ has the
following form,
\begin{equation}
A=R_{z}(\psi)R_{y}(\theta)R_{x}(\varphi)\\
=\left(%
\begin{array}{ccc}
  \cos\theta\cos\psi & \cos\theta\sin\psi & -\sin\theta \\
  \sin\varphi\sin\theta\cos\psi-\cos\varphi\sin\psi & \sin\varphi\sin\theta\sin\psi+\cos\varphi\cos\psi& \sin\varphi\cos\theta \\
  \cos\varphi\sin\theta\cos\psi+\sin\varphi\sin\psi  & \cos\varphi\sin\theta\sin\psi-\sin\varphi\cos\psi & \cos\varphi\cos\theta \\
\end{array}%
\right),
\end{equation}
where $R_{x}(\psi)$, $R_{y}(\theta)$, and $R_{z}(\varphi)$ stand
for the rotation along the x-axis, y-axis, and z-axis of the
global coordinate system, respectively.
We assume the direction of the grain polarization is along the
direction of $x_{3}^{'}$ of the grain in the local coordinate
system. By random selecting three angles $\psi$, $\theta$, and
$\varphi$ between $0$ and $2\pi$ for each grain in the global
coordinate system, we can obtain a random distribution of the
polarizations in the ceramics. The averaged polarization along the
$x_{3}$ direction in the global coordinate system is selected as
the measurable polarization of the ceramics. The ceramic is
simulated by a cylindrical lattice, each lattice site is
distributed with a grain. The radial and height of the cylinder
lattice are $15$ and $60$, respectively.

  The main motivation of this paper is to give an explanation on the
path dependence electromechanical poling process mentioned in
Ref.~6. In the experiment, two poling protocols are designed as
following: For protocol $1$, the poling field is raised on the
sample from zero to a maximum value $E_{max}$ at first, then a
radial stress is loaded, and the mechanical stress is removed
after a period of time, then the poling field is turned down from
the maximum value to zero. For protocol $2$, the radial stress is
loaded on the sample at first, then a poling field cycle is
applied on the sample before the stress is removed. During the
above two protocols, the domain switching is induced either by the
electric field or by the combined electric and mechanical loads.
For the first case, i.e., only under an electric load,
 the $180^{0}$  switching occurs when the total work done reaches
a critical value($E_{180}$) corresponding to the energy barrier
 of $180^{0}$ domains, and the $90^{0}$ domain switching occurs when
the total work done reaches a critical value($E_{90}$)
corresponding to the energy barrier
 of $90^{0}$ domains.
If only under the application of electric field, the switching of
$180^{0}$ domain can only be $180^{0}$ switching, and the
switching of $90^{0}$ domain can only be $90^{0}$ switching.
Mathematically, the domain switching criterions for $90^{0}$ and
$180^{0}$ switchings under an electric load can be given by
\begin{equation}
\begin{array}{c}
E_{i}\cdot\Delta P_{i} \geq E_{90},  \\
E_{i}\cdot\Delta P_{i} \geq E_{180}. \\
\end{array}
\end{equation}

For the second case, i.e., the case that domain-switching is
driven by both electric load and mechanical load. A
two-step-switching criterion~\cite{two-step1,two-step2} is used
here to simulate the different experimental results between
protocol 1 and protocol 2. As referred in Ref.~6 , the switching
criterion of Eq.~(1) cannot obtain the path dependence of the
remnant polarization for different poling protocols. The
two-step-switching model divides each $180^{0}$ switching into two
successive $90^{0}$ switchings, and the $90^{0}$ switching
criterion is described as following:
\begin {equation}
E_{i}\cdot\Delta P^{s}_{i}+\sigma_{ij}\cdot\Delta e_{ij}^{s}\geq
E_{90}.
\end{equation}
In our simulation, the mechanical stress is applied along the
radial direction, the electric filed is applied along the
z-direction in the global coordinate system, we resolve the stress
and the spontaneous strain into x-direction and y-direction, and
the work done by the stress on each domain $i$ during the
switching process is calculated as the summation $\sigma_{x}\Delta
e_{ix}^{s}+\sigma_{y}\Delta e_{iy}^{s}$.
The spontaneous strain of the tetragonal unit cell has principal
values $e_{3}=(c-a_{0})/a_{0}$ and $e_{1}=e_{2}=(a-a_{0})/a_{0}$,
where $a$ and $c$ are the lattice parameters of the tetragonal
unit cell and $a_{0}$ is the lattice parameter of the cubic cell
above the Curie temperature. We assume
$e_{1}=e_{2}=-\frac{1}{2}e_{3}$ with $e_{3}=e_{0}$ within the
whole paper. From Eq.~(4), we can see that the lateral stress will
hinder the first $90^{0}$ switching if the strength of electric
field is not strong enough, especially for these grains whose
polarizations are almost antiparallel to the external electric
field.
 In order to take the
interaction of the grain with its surrounding grains into account,
the inclusion model is introduced into our
simulation~\cite{Hwang2}. Therefore, the domain switching
criterion for the ceramics under both electric load and mechanical
load is modified as following:
 \begin {equation}
(E_{i}+\frac{1}{3\bar{\epsilon}}P_{i}^{sm})\cdot\Delta
P^{s}_{i}+(\sigma_{ij}+\frac{2}{5}\bar{Y}e_{ij}^{sm})\cdot\Delta
e_{ij}^{s}\geq E_{90},
\end{equation}
and the domain switching criterion in Eq.~(3) for the ceramics
only under an electric load is changed into

\begin{equation}
\begin{array}{c}
E_{i}+\frac{1}{3\bar{\epsilon}}P_{i}^{sm})\cdot\Delta
P^{s}_{i}\geq E_{90},\\
(E_{i}+\frac{1}{3\bar{\epsilon}}P_{i}^{sm})\cdot\Delta
P^{s}_{i}\geq E_{180}, \\
\end{array}
\end{equation}

where $\bar{\epsilon}$ is the effective dielectric permittivity of
the ceramics, $\bar{Y}$ is the effective modulus of the ceramics,
$P_{i}^{sm}$ is the averaged spontaneous polarization of the
matrix of grain $i$, and $e_{ij}^{sm}$ the averaged spontaneous
strain of the matrix of grain $i$.

 We use a cylindrical lattice with radial $15$ and height $60$ to simulate
the cylindrical ceramics sample in the experiment~\cite{Exp}, each
lattice site is distributed with a grain. The polarization of the
grain is selected randomly by random choosing three angles $\psi$,
$\theta$, and $\varphi$ for each grain. So, the averaged total
polarization of the sample is zero at initial time.  The volume
fraction of $90^{0}$ domain is taken to be $0.1$, which can be
realized by randomly selecting $10\%$ sites, and do a mark for
these polarizations. For $90^{0}$ domain, only $90^{0}$ switching
can be taken place under an electric load. For $180^{0}$ domain,
only the $180^{0}$ switching is permitted under an electric load.
When the sample is applied by an electric field and a mechanical
stress simultaneously, the two-step-switching model is adopted for
both the $90^{0}$ domains and $180^{0}$ domains. It is assumed
that the relax time of domain switching is very short compared
with the change of the electric field and mechanical stress. In
protocol $1$ and protocol $2$, when the sample is only under the
electric load, the criterion described in Eq.~(6) is used to
decide whether a $90^{0}$ switching or a $180^{0}$ switching is
taken place. So, for $180^{0}$ domain, we can try to reverse the
polarization of the grain in the local coordinate system, the
direction of the grain polarization will take the new direction
when the second formula in Eq.~(6) is satisfied, otherwise, the
direction of the grain polarization keeps invariant. For $90^{0}$
domain, there are four possible new directions in the local
coordinate system, we can try to re-orientate the polarization to
one of the new directions, and if the new direction meets the
first criterion described in Eq.~(6), then the domain rotates to
the new direction, if two or more of the new directions meet the
criterion simultaneously, the domain switches to the one that
makes the left hand side of the first formula in Eq.~(6) the
largest. When the sample is under combined electric and mechanical
loads, the two-step-switching is simulated as following: the first
$90^{0}$ switching is determined among four possible new
directions dependent on whether the condition in Eq.~(5) is
satisfied. Only the first $90^{0}$ switching occurs realistically,
should we try to decide whether the second $90^{0}$ switching can
occur. The possible test direction for the second $90^{0}$
switching should be the inverse direction of the grain before the
first $90^{0}$ switching. The simulation process is carried out
from a randomly orientated initial state, and then the electric
field is increased from zero to $750$V/mm with a step length
$15$V/mm. At each step of the applied electric field, a scan over
all the grains is executed, and the polarization direction of each
grain is simultaneously decided whether to take a new direction to
make the ceramics staying at a low energy state. The total
polarization of the ceramics is calculated at each electric field.
After the increasing process, the applied electric field is then
decreased from 750V/mm to zero. The parameters for
$Pb_{0.99}[Zr_{0.45}Ti_{0.47}(Ni_{0.33}Sb_{0.67})_{0.08}]O_{3}$
used in our simulation is taken from Ref.~5,8 and are as
following: $P_{s}=0.45C/m^{2}$, $e_{0}=0.0073$, $E_{90}=0.13$MV/m,
$E_{180}=2E_{90}$, $\bar{\epsilon}=0.80\mu$ F/m, $\bar{Y}=7.5$Gpa,
the effective stress loaded on the ceramics along x-direction and
y-direction is selected as $20$Mpa to fit
the experiment data.\\
\section{Results and Discussions}

Our simulation results on the polycrystalline ferroelectric
ceramics are shown in Fig.~2, where triangle symbols and circle
symbols denote the experiment data in protocol $1$ and protocol
$2$, respectively, and solid lines and dotted lines stand for our
simulation results in protocol $1$ and protocol $2$, respectively.

\begin{figure}
\includegraphics[width=4.0in]{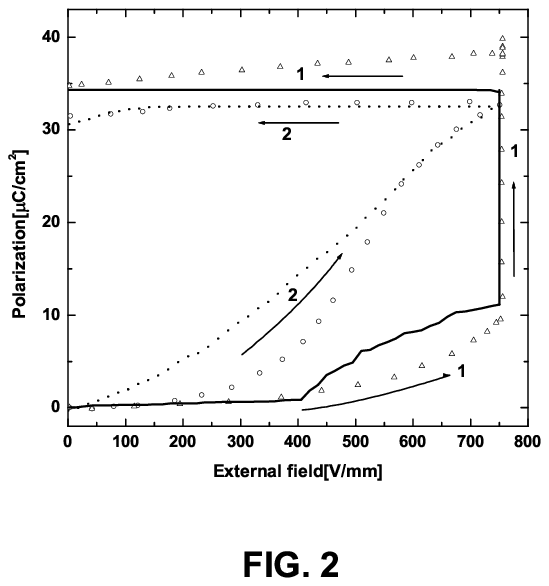}
\end{figure}
From Fig.~2, we can see that there exists a clear difference of
the remnant polarization between protocol $1$ and protocol $2$,
which cannot be given by Hwang's switching criterion. Furthermore,
the difference of the remnant polarization between the two
protocols will increase with the increase of radial stress loaded
perpendicular to the electric field. The path dependent remnant
polarization is aroused by the first $90^{0}$ switching of these
polarizations which are almost antiparallel to the electric field.
When the radial stress is loaded, the first $90^{0}$ switching of
the above mentioned polarizations are difficult to occur. From
Eq.~(5), we can see that the larger the stress is, the more
difficult the first $90^{0}$ switching occurs for these
polarizations whose directions are close to the inverse direction
of electric field. If the electric field is loaded before the
application of lateral stress, these polarizations, whose
directions are close to the inverse direction of electric field,
are already switched under the action of electric field when the
switching criterion Eq~.(6) is satisfied. It is the reason why the
remnant polarizations are different between the two protocols. If
only the x-direction lateral stress in the global coordinate
system is loaded for a cubic sample~\cite{two-step2}, we can
choose a specific grain, which direction is exactly inverse to the
direction of the electric field, to illustrate that there is no
difference in remnant polarization between the two cases that the
mechanical stress is applied before and after the application of
electric field. The polarization of the above grain will first
flip from minus z-direction to y-direction, then flip from
y-direction to z-direction, and the stress in this case won't
hinder the switching of the polarizations. As a result, the
averaged polarizations along z-axis will be the same for the two
protocols. So, the radial stress is also an important factor for
the occurrence of the difference of the remnant polarization
between the two protocols. The small deviation of our simulation
results with the experiment data in Fig.~2 may be resulted from
the inclusion model, which does not take the shape of the sample
into account. Further investigation is currently in progress to
take the neighbor dipole-dipole interactions into account rather
than the inclusion model. Despite the simulation being carried out
with the assumption that the PZT is tetragonal, the model is still
physically justifiable for the rhombohedral material, and the
$90^{0}$ tetragonal switches can provide a reasonable
representation of the $\sim70^{0}$ and $\sim110^{0}$ switches in
the rhombohedral phase.

In summary, we employ the two-step-switching model to simulate the
electromechanical poling behavior of unpolable ferroelectric
ceramics. It is found that the application of mechanical stress
perpendicular to the electric poling direction will increase the
remnant polarization, especially when the radial mechanical stress
is loaded after the application of electric field. An explanation
on the path dependence of the polarization buildup is presented.

\begin{acknowledgments}
This work was partially supported by the Science Foundation of the
education committee of Jiangsu province under the grant
No.~07KJB140002, and the open project of Jiangsu Lab of Advanced
Functional Material under the grant No.~06KFJJ007.
\end{acknowledgments}

\newpage
 Figure Captions\\

FIG1: Schematic diagram of the electric-mechanical loading
system.\\

FIG2: Polarization of the ferroelectric ceramics as a function of
poling field for different protocols.
   Solid line and dotted line are our simulation results for
   protocol 1 and protocol 2, respectively. Triangle symbols
   and circle symbols denote the experimental data~\cite{Exp} for the two
   protocols.
\newpage

\end{document}